\documentclass{optica-article}

\journal{opticajournal} 

\articletype{Research Article}

\usepackage{lineno}

\usepackage{empheq}
\usepackage{textcomp}
\usepackage{bigints}
\usepackage{colortbl}
\usepackage{multirow}
\usepackage{array}
\usepackage{graphicx}
\usepackage{braket}
\usepackage{float}
\usepackage{bigints}
\usepackage{makecell}
\usepackage{graphicx}
\usepackage{dcolumn}
\usepackage{bm}
\usepackage{hyperref}
\hypersetup{
    colorlinks = true,
    linkcolor = black,
    citecolor=blue, 
    }
\usepackage[capitalise]{cleveref}
\usepackage{siunitx}
\usepackage{graphicx} 
\usepackage{booktabs}
\usepackage{chemformula}

\begin{document}

\title{Hybridized-band parametric oscillations in coupled Kerr microresonators}

\author{Luca O. Trinchão,\authormark{1}
Luiz Peres,\authormark{1}
Eduardo S. Gonçalves,\authormark{1}
Miguel Nienstedt,\authormark{1}
Laís Fujii dos Santos,\authormark{2}
Paulo F. Jarschel,\authormark{1}
Thiago P. Mayer Alegre,\authormark{1}
Nathalia B. Tomazio,\authormark{3,4}
and Gustavo S. Wiederhecker,\authormark{1,*}
}

\address{\authormark{1}Gleb Wataghin Institute of Physics, University of Campinas, Campinas, SP, Brazil\\
\authormark{2}School of Electrical Engineering and Computer Science, University of Ottawa, Ottawa, ON, Canada\\
\authormark{3}Instituto de Física, Universidade de São Paulo, São Paulo, SP, Brazil\\
\authormark{4}Leibniz Institute of Photonic Technology, Jena, Germany}

\email{\authormark{*}gsw@unicamp.br} 


\begin{abstract*} 
Coupled resonators form band-like optical states that support rich nonlinearities beyond what is possible in single resonators.
In these systems, four-wave mixing mediates interband coupling, displaying multimode dynamics that span both spatial and spectral degrees of freedom.
In this study, we propose a framework describing the onset and control of hybridized optical parametric oscillation in three coupled silicon nitride microring resonators.
In a symmetric configuration, we observe the emergence of diverse phase-matching pathways defined by the dispersive band structure.
We develop an analytical model that captures the parametric gain of these interband processes and derive closed-form expressions for the dominant gain maxima; the analytical framework itself readily extends to more complex coupled networks.
We further report an asymmetric design that co-engineers mode overlap and dispersion to operate on a compact 7-GHz spacing, free from mode competition.
Our findings establish design principles for engineering nonlinear dynamics in coupled-resonator platforms, with implications for coherent photonic computing and quantum information processing.
\end{abstract*}

\section{Introduction}
Multimode photonics has become an exciting field for exploring interaction pathways otherwise unattainable in single-mode devices.
Examples include the use of different transverse mode families in waveguides and resonators~\cite{mohanty2017quantum, feng2016chip, amorim2025spatial}, interplay between axial and azimuthal modes in microcavities~\cite{sumetsky2011surface, eadie2025azimuthal}, cross-polarized interactions in fibers and waveguides~\cite{rodrigues2025cross, rehan2025second, quinn2024coherent, quinn2023random}, and hybridized states in topological systems~\cite{tomazio2024tunable, trinchão2025mapping, flower2024observation, mazanov2024photonic, roberts2022topological, de2023heterogeneously}. 
The latter has been implemented in coupled-resonator platforms, enabling notable work in neuromorphic computing~\cite{wanjura2024fully, biasi2024exploring} and symmetry-breaking of light~\cite{pal2024linear, ghosh2024controlled, ghosh2025spontaneous}.

\begin{figure*}
\centering
\includegraphics[width=\textwidth]{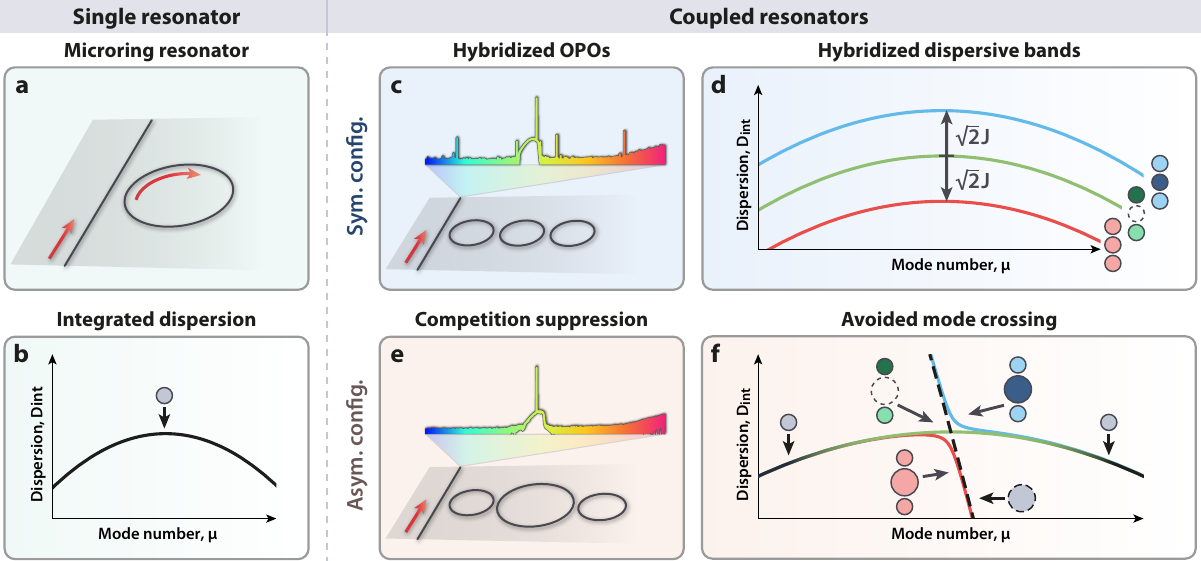}
\caption{\label{fig:1}
\textbf{(a)} Schematic of a microring resonator driven by a monochromatic pump (red arrow).
\textbf{(b)} Representative integrated dispersion of a single resonator operating in the normal-dispersion regime.
\textbf{(c)} Generation of hybridized optical parametric oscillations (OPOs) in a symmetric three-ring resonator under monochromatic pumping (red arrow).
\textbf{(d)} Dispersive band structure of the $|$ooo$|$ design, showing the S (red), C (green), and AS (blue) supermode bands.
The associated spatial profiles are also shown: dark shading indicates a relative $\pi$ phase shift, while blank rings denote non-resonant excitation.
\textbf{(c)} Hybridized OPO operation in an asymmetric three-ring resonator formed by larger and smaller rings, in which geometric asymmetry suppresses competing comb-generation channels.
\textbf{(d)} Dispersive band structure of the $|$oOo$|$ design, highlighting an avoided mode crossing between the dispersions of the smaller and larger rings (black dashed and solid lines). The transition of the supermode eigenvectors from the strongly coupled (colored) to the uncoupled regime (black solid) is also shown.
}
\end{figure*}

Coupled microresonators operate in the strong-coupling regime, where near-field evanescent coupling produces frequency-split hybridized supermodes~\cite{haus2002coupled, tomazio2024tunable, trinchão2025mapping}. 
These supermodes reshape both spectral and spatial field features, reflected by the eigenfrequencies and eigenvectors of the coupled system~\cite{tomazio2024tunable, trinchão2025mapping}, and collectively form a multimode dispersive band structure~\cite{tikan2021emergent}.
The band multiplicity is determined by the number of supported supermodes, while their topology is shaped by the combined contributions of the bare-ring group velocity dispersion $D_{\mathrm{int}}$ and the inter-ring coupling $J$, as illustrated in \cref{fig:1}(c,d) for three-ring configurations.
This spectral branching unlocks new four-wave mixing (FWM) pathways through hybridized interband interactions, describing photon transfer across the supermodes of the system~\cite{tikan2021emergent, ji2025multicolor, gao2024observation, sanyal2025nonlinear}.

Despite recent investigations of coupled microresonators in the context of Kerr frequency combs~\cite{ji2025multicolor, gao2024observation, tikan2021emergent, sanyal2025nonlinear}, their behavior under individual $\chi^{(3)}$-based (hyperparametric) interactions has received comparatively less attention~\cite{zeng2014design}.
For single resonators, a comprehensive framework has been established for the modulation instability underlying Kerr comb formation, yielding closed-form expressions for the primary comb sidebands that depend primarily on the cavity dispersion~\cite{herr2012universal}.
In coupled-resonator platforms, theoretical advance has revealed a range of novel nonlinear phenomena~\cite{tikan2021emergent}, including Kelly sidebands~\cite{gao2024observation} and the Moiré speedup effect~\cite{ji2024multimodality}.
However, analytical treatments of the onset of interband FWM that explicitly capture the role of the inter-ring coupling function $J(\mu)$, or that incorporate the distinct spatial profiles of the supermodes~\cite{trinchão2025mapping, zeng2014design}, remain comparatively limited.
Beyond nonlinear dynamics, coupled resonators have also attracted significant interest for quantum information processing.
On the one hand, their high mode density readily supports phase-matched four-wave mixing~\cite{mehrabad2025multi}, but often also leads to mode competition~\cite{tomazio2024tunable}, resulting in parasitic noise that can degrade the purity of squeezed states~\cite{zhang2021squeezed, ulanov2025quadrature, sabattoli2021suppression, seifoory2022degenerate, zhao2020near}.
On the other hand, interband interactions are intrinsically multimode and provide a natural pathway toward spatial–frequency hyperentanglement~\cite{deng2017quantum, chiriano2023hyper}.

In this work, we introduce a supermode framework for Kerr-mediated interband interactions in coupled resonator systems and apply it to hybridized optical parametric oscillations (OPOs) in a symmetric three-ring molecule ($|$ooo$|$, \cref{fig:1}(c,d)).
Hyperparametric oscillations are considered, wherein degenerate FWM produces correlated signal and idler fields at equidistant frequencies from the monochromatic pump ($2 \,\omega_\mathrm{pump}\to \omega_\mathrm{signal} + \omega_\mathrm{idler}$)~\cite{lu2025photonic}.
We experimentally identify three distinct excitation pathways, in excellent agreement with coupled-mode theory for interband parametric gain.
These hybridized oscillations can be understood with a set of nonlinear coupled supermode equations (see Supplementary Material Section S2).
In this picture, each supermode is a potential participant in the FWM process, with dynamics shaped by supermode-dependent overlaps and phase-matching.
Using this formalism, we derive closed-form expressions for the first FWM sidebands associated with distinct phase-matching mechanisms in a three-resonator platform.
However, the methodology readily extends to higher-order networks, requiring only knowledge of the coupling topology~\cite{de2023coupled, zeng2014design}, the coupling function, and the group velocity dispersion of the individual resonators.
Guided by this framework, we introduce an asymmetric three-ring design ($|$oOo$|$, \cref{fig:1}(e,f)) that co-engineers these conditions via avoided-mode crossing (AMX).
We demonstrate that geometric asymmetry quenches parasitic channels, enabling a competition-free OPO with 7-GHz signal–idler spacing, within the bandwidth of standard photodetectors.
Overall, our results establish practical design strategies for engineering nonlinear interactions in coupled resonator platforms, with direct implications for photonic coherent computing and quantum information processing.

\section{Dispersive band structure}

We study the hybridized interactions of a three-ring coupled system (\cref{fig:1}(c-f)).
Evanescent coupling induces the formation of three orthogonal supermode eigenstates~\cite{tomazio2024tunable, trinchão2025mapping}, referred to as symmetric (S), central (C), and anti-symmetric (AS), with $\omega_\mathrm{S}<\omega_\mathrm{C}<\omega_\mathrm{AS}$ (see Supplemental Material S1). 
In this system, the effective dispersion landscape is shaped by the underlying bare-ring group velocity dispersion, $D_{\mathrm{int}}^{{(0)}}(\mu)$, and dispersive inter-ring coupling, $J(\mu)$. For identical (degenerate, $|$ooo$|$) rings, the resulting supermode integrated dispersions are given by:
\begin{equation}
    \begin{split}
                D_{\mathrm{int}}^{(\mathrm{S})}(\mu)
            &= \omega^{(\mathrm{S})}(\mu) - \omega^{(\mathrm{C})}_{0} - D_1^{(\mathrm{C})} \mu \\
            &= D_{\mathrm{int}}^{(\mathrm{C})}(\mu) - J(\mu)
    \end{split}
\end{equation}
and
\begin{equation}
    \begin{split}
            D_{\mathrm{int}}^{(\mathrm{AS})}(\mu)
            &= \omega^{(\mathrm{AS})}(\mu) - \omega^{(\mathrm{C})}_{0} - D_1^{(\mathrm{C})} \mu \\
            &= D_{\mathrm{int}}^{(\mathrm{C})}(\mu) + J(\mu),
    \end{split}
\end{equation}
where $\omega^{(\mathrm{C})}_{0}$ is the frequency of the pumped C supermode at $\mu = 0$, and $D_1^{(\mathrm{C})}$ is its free spectral range (FSR). The C supermode inherits the bare-ring dispersion:
\begin{equation}
    D_{\mathrm{int}}^{(\mathrm{C})}(\mu) = D_{\mathrm{int}}^{(0)}(\mu).
\end{equation}
\cref{fig:1}(d) shows the typical dispersive band structure for the S, C, and AS supermodes under normal dispersion of the bare ring. The case of asymmetric rings ($|$oOo$|$, \cref{fig:1}(e,f)), where dispersion mismatch of the bare rings results in extra modifications of the dispersive bands through AMX, will be discussed in \cref{sec:competition_suppression}.

\begin{figure}
    \centering
    \includegraphics[width=.65\linewidth]{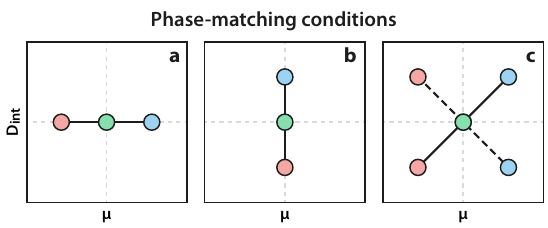}
    \caption{Phase-matching topologies of hybridized OPOs.
Phase matching in coupled microresonators can be classified according to the geometry of the interaction within the dispersive band structure. (a) Horizontal phase matching, corresponding to a Type-I-like OPO, where signal and idler are generated on the same supermode branch (AS–AS) at finite azimuthal order through an effective anomalous dispersion induced by band curvature. (b) Vertical phase matching, corresponding to a Type-II-like OPO at $\Delta\mu = 0$, where signal and idler occupy different branches (S and AS) and frequency matching is enabled solely by interband splitting. (c) Diagonal phase matching, also Type-II-like, where signal and idler are generated on different branches at finite and opposite azimuthal orders, requiring simultaneous interband and intraband dispersion compensation. While panels (b) and (c) share the same modal degeneracy class, they represent distinct dynamical regimes with different gain maxima and competition behavior. In (a-c), the color coding indicates the frequency hierarchy of the interaction
    }
    \label{fig:2}
\end{figure}

In single-ring resonators, FWM phase-matching is typically restricted to multi-FSR interactions, where anomalous dispersion and nonlinearity counterbalance each other~\cite{herr2012universal, inga2020alumina, fujii2020dispersion, soares2023third}. 
Coupled resonators, by contrast, introduce \emph{verticality} to the phase-matching condition via interband interactions, enabled by the formation of hybridized supermode dispersion branches. 
This additional flexibility enables the realization and tailoring of FWM interactions that are otherwise unattainable in single resonators.
As illustrated in \cref{fig:2}, hybridized optical parametric oscillations (OPOs) can be systematically classified using two independent descriptors: the \emph{modal degeneracy class}, which specifies how signal and idler are distributed among supermode branches, and the \emph{phase-matching topology}, which describes how frequency and momentum conservation are satisfied within the dispersive band structure. 
By analogy with the conventional Type-0/Type-I/Type-II nomenclature used in $\chi^{(2)}$ parametric processes, we define Type-II-like OPOs as those in which signal and idler occupy different supermode branches (e.g., S and AS), and Type-I-like OPOs as those in which signal and idler reside on the same branch distinct from the pump (e.g., pump at C, signal and idler at AS branches). In this generalized usage, the supermode branch plays the role customarily associated with polarization, and the nomenclature refers exclusively to modal degeneracy rather than to the order of the nonlinearity.

Within this framework, OPOs may arise through distinct phase-matching topologies. 
First, horizontal phase matching (\cref{fig:2}(a)) corresponds to a Type-I-like process, where signal and idler are generated on the same supermode branch at finite azimuthal order through an effectively dispersiveless or quasi-anomalous frequency grid defined by $D_{\mathrm{int}} \approx 0$. This pathway corresponds to the typical route observed in single-resonator geometries at later stages of evolution, after XPM has taken effect~\cite{herr2012universal}. 
Second, vertical phase matching (\cref{fig:2}(b)) corresponds to a Type-II-like interband OPO occurring at a single azimuthal order $\mu$ ($\Delta\mu = 0$), where frequency matching is satisfied purely by the supermode splitting and is unique to coupled-resonator platforms~\cite{tomazio2024tunable}. 
Third, diagonal phase matching (\cref{fig:2}(c)) also corresponds to a Type-II-like process, but involves finite and opposite azimuthal orders ($\Delta\mu \neq 0$), requiring simultaneous interband and intraband dispersion compensation; this regime has recently been demonstrated in the context of multicolor solitons~\cite{ji2025multicolor}. 
Although vertical and diagonal pathways share the same modal degeneracy class, they represent distinct dynamical regimes with different gain spectra and competition behavior. We note that the labels OPO~1 and OPO~2, introduced in the following section, refer solely to the order in which the oscillations are observed experimentally and do not imply a correspondence with the Type-I / Type-II-like classification.

Across all pathway scenarios, the frequency spacing follows an equally spaced rule, with horizontal interactions spanning hundreds of gigahertz and vertical interactions only a few.
Moreover, interband interactions are less constrained by dispersion and have been demonstrated in the normal regime~\cite{tomazio2024tunable}.
We further acknowledge the capability of interband interactions in Bragg-grating cavities via coupling of counterpropagating modes~\cite{moille2023fourier, stone2024wavelength, kheyri2025chip, ulanov2025quadrature}. 
However, these platforms typically lack the post-fabrication tunability available in coupled microresonators~\cite{tomazio2024tunable, gentry2014tunable, pidgayko2023voltage, xia2025reconfigurable}.

\section{Hybridized OPOs}

\begin{figure}
\centering
\includegraphics[width=\textwidth]{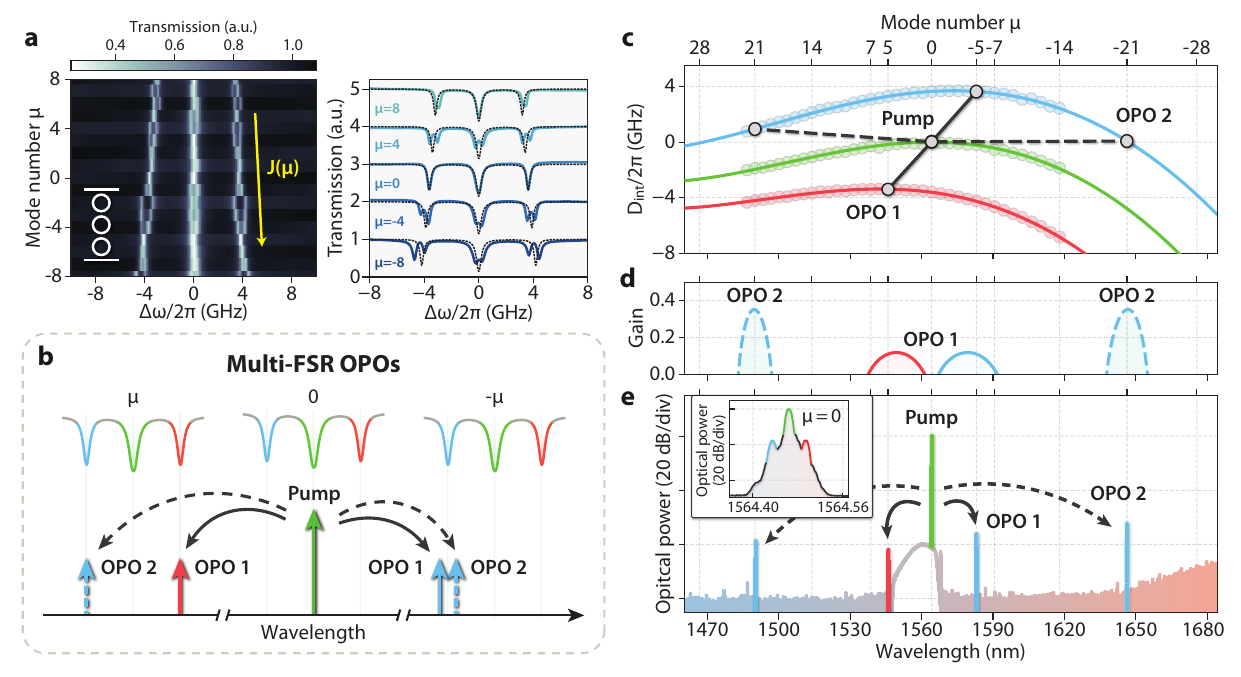}
\caption{\label{fig:3}
Hybridized OPOs in the $|$ooo$|$ coupled resonator system.
\textbf{(a)} Left: Transmission map as a function of frequency offset ($\Delta\omega/2\pi$) and relative mode number $\mu$. 
Each FSR supports a hybridized supermode triplet, with splitting increasing at longer wavelengths due to stronger coupling.
Additional splittings are visible at some $\mu$ due to CW–CCW backscattering.
Right: Transmission spectra for selected $\mu$ values. Dashed lines represent the theoretical transmission.
\textbf{(b)} Schematic of the two observed multi-FSR OPOs. 
Pumping the C supermode excites distinct oscillations: OPO 1 (solid) generates signal and idler in the S and AS supermodes, while OPO 2 (dashed) produces both within AS supermodes.
\textbf{(c)} Integrated dispersion of the supermodes, referenced to the pumped C mode at $\mu = 0$. 
Colored markers represent experimental data, while solid curves show polynomial fits. 
Modes participating in OPO 1 and OPO 2 are highlighted by solid and dashed black lines, corresponding to diagonal and horizontal phase-matching, respectively.
\textbf{(d)} Calculated parametric gain spectra. For OPO 1 (solid), gain peaks for the S and AS modes stem near $\mu = \pm 4$. For OPO 2 (dashed), the AS–AS gain peaks arise at $\mu = \pm 21$.
\textbf{(e)} Optical spectrum showing simultaneous excitation of OPO 1 and OPO 2. Inset: zoom at $\mu = 0$, revealing an intra-FSR OPO enabled by vertical phase-matching.
}
\end{figure}

In the symmetric three-ring system ($|$ooo$|$), evanescent coupling induces hybridized triplets across the entire optical spectrum~\cite{trinchão2025mapping}. \cref{fig:3}(a) shows the transmission spectrum of several triplets, indexed by the relative azimuthal mode number $\mu = m - m_0$, where $m_0$ corresponds to the mode near \qty{1564.4}{\nano\meter}. As the optical wavelength increases (i.e., $\mu$ decreases), the evanescent tail penetrates further into the surrounding rings, increasing the coupling strength.
This results in a larger frequency splitting between the supermodes over the relevant wavelength range~\cite{povinelli2005evanescent}. For some specific $\mu$ values, additional mode splitting is observed due to backscattering-induced coupling between the clockwise (CW) and counterclockwise (CCW) propagating modes.

\begin{figure}[ht]
    \centering
    \includegraphics[width=.5\linewidth]{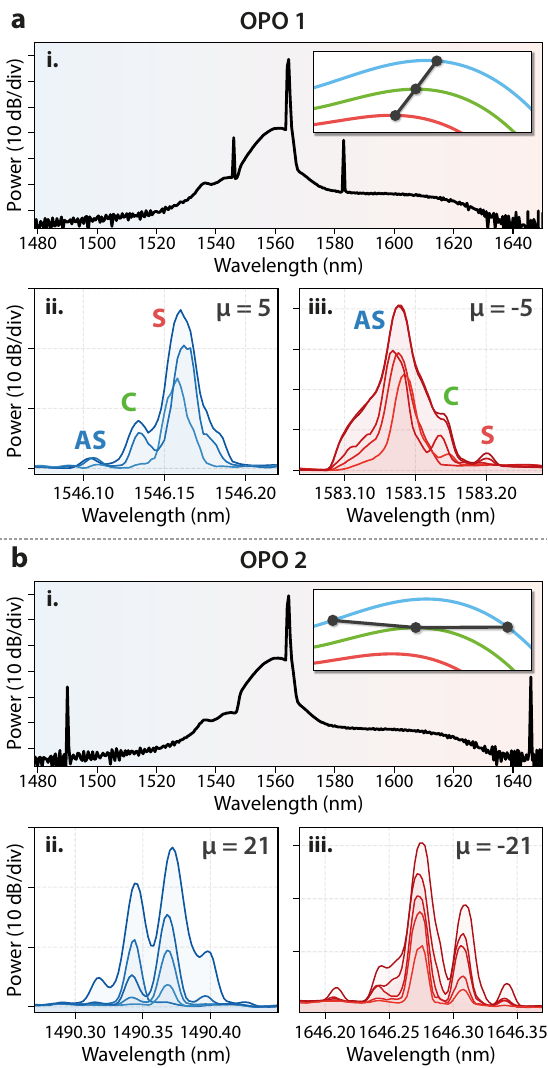}
    \caption{
    OPO 1 \textbf{(a)} and OPO 2 \textbf{(b)} under individual excitation.
    \textbf{(i)} Broadband optical spectrum, with insets showing the corresponding phase-matching pathways.
    \textbf{(ii–iii)} Zoom into the signal and idler frequencies at $\mu = \pm 5$ for (a) and $\mu = \pm 21$ for (b), with incremental adjustments of pump power and detuning to track their evolution.
    In \textbf{(a, ii–iii)}, the excited supermodes are identified.
    In \textbf{(b, ii–iii)}, additional lines are observed.
    }
    \label{fig:4}
\end{figure}

Experimental investigation of hybridized OPOs is carried out in a commercial \ch{Si3N4} platform cladded in silica. An arrangement of three \qty{50}{\micro\meter}-radius ring resonators is coherently driven using an external tunable continuous-wave diode laser. 
We compensate for the cavity's large mode volume and moderate quality factors ($Q^{(C)}_L = 4 \times10^5$, loaded) by modulating the pump into optical pulses (\qty{250}{\nano\second} width, \qty{10}{\micro\second} period) using an electro-optic modulator, followed by amplification with an erbium-doped fiber amplifier.
The peak powers (in the waveguide) used for the $|$ooo$|$ (\cref{fig:3,fig:4}) and $|$oOo$|$ (\cref{fig:5}) measurements were \qty{2.01}{\watt} and \qty{1.67}{\watt}, respectively.
Moreover, by operating each ring individually in the normal dispersion regime, which would otherwise suppress phase-matching for single-ring OPOs, we isolate and emphasize the interband interactions, which are the only processes supported in this device.

\cref{fig:3}(b) illustrates the two dominating OPO processes observed when pumping a C supermode in the $|$ooo$|$ design. The first process, denoted OPO~1, corresponds to a \emph{diagonal Type-II-like interband OPO}, in which the signal and idler are generated in different supermode branches (AS and S, respectively) at finite and opposite azimuthal indices.
The second process, denoted OPO~2, corresponds to a \emph{horizontal Type-I-like hybridized-band OPO}, where both signal and idler emerge within the same AS supermode branch.
We approach both processes theoretically and experimentally below.

\cref{fig:3}(c) shows the measured dispersive bands for the S, C, and AS supermodes referenced with respect to the pumped C mode at $\mu = 0$. 
The modes participating in OPO 1 and 2 are highlighted. 
In OPO 1 (solid line), diagonal phase-matching is achieved when the frequency mismatch of the S and AS branches simultaneously vanishes. These mismatches are expressed as
\begin{align}
    \delta_\mu^{(\mathrm{S})}(\mu) &=  D^{(0)}_{int}(\mu) - \sqrt{2}J(\mu) + \Delta_p - 2\Gamma_{\mathrm{S},\mathrm{C}}\bar{P}, 
    \label{eq:freq_match_S}\\
    \delta_\mu^{(\mathrm{AS})}(\mu) &=  D^{(0)}_{int}(\mu) + \sqrt{2}J(\mu) + \Delta_p - 2\Gamma_{\mathrm{AS},\mathrm{C}}\bar{P}
    \label{eq:freq_match_AS}
\end{align}
where $\Delta_p = \omega^{(\mathrm{C})}_0 - \omega_p$ is the pump detuning. The final terms on the RHS represent the Kerr-induced XPM from the pumped C mode, with $\Gamma_{\mathrm{S},\mathrm{C}}$ and $\Gamma_{\mathrm{AS},\mathrm{C}}$ being the corresponding supermode spatial overlaps~\cite{trinchão2025mapping,tomazio2024tunable} and $\bar{P}$ the normalized intracavity pump power (see Supplementary Material Section S2).

\cref{eq:freq_match_S,eq:freq_match_AS} show that, in coupled-resonator platforms, efficient phase-matching requires balancing out the bare-ring dispersion, the inter-ring coupling, the pump detuning, and the nonlinear XPM-induced frequency shift.
The first (primary) OPO sidebands arise at azimuthal order $\mu_{th,_I}$ corresponding to maximum parametric gain. For OPO 1, this threshold admits a closed-form expression (see Supplementary Material, Section S2),
\begin{equation}
    \mu_{th,I}=\frac{\sqrt{2}J_1}{D_2^{(0)}},
\end{equation}
which depends exclusively on the first-order derivative of the coupling function, $J_1=dJ/d\mu$, and the bare-ring dispersion.
This result shows that the coupling function formula $J(\mu)$ plays a role as important as the cavity dispersion in shaping nonlinear interactions in coupled-resonators.
In fact, the position of the primary sidebands can therefore be tuned via engineering the inter-ring coupling, for instance by adjusting the resonator gap.
In \cref{fig:3}(d), we plot the calculated parametric gain lobes, as derived in the Supplementary Material Section S2.
Experimental verification in \cref{fig:3}(e) shows the OPO 1 sidebands being excited at $\mu = \pm 5$, in excellent agreement with the predicted gain maxima.

Moreover, OPO 2 (dashed) describes signal and idler generation exclusively within the AS supermode family, with both sidebands residing on the same dispersive branch.
This process occurs when the AS-branch integrated dispersion $D_{\mathrm{int}}^{(\mathrm{AS})}(\mu)$ is slightly positive and close to zero at $\pm\mu_{th,II}$, forming an effectively quasi-horizontal dispersion regime.
This can be understood as the interband interaction coupling two AS modes to the pumped C supermode, such that the three participating modes define a positively concave dispersion parabola (see dashed black line in \cref{fig:3}(c)).
As a result, OPO 2 experiences an \emph{effective} anomalous dispersion that emerges from the topology of the hybridized bands, rather than from the inherent normal dispersion of the individual resonators.

In this regime, the threshold azimuthal order for the OPO 2 process also admits a closed-form expression,
\begin{equation}
\mu_{th,II} =
    \sqrt{\frac{\bar{P} - 2\sqrt{2}J_0 -2\Delta_p}{D_2^{(0)}+\sqrt{2}J_2}},
\end{equation}
with $J_0=J(\mu=0)$ and $J_2=d^2J/d\mu^2$.
For the parameters of our device, this condition is met near $\mu = \pm 21$, in agreement with both the calculated parametric gain shown in \cref{fig:3}(d) and the experimental optical spectrum in \cref{fig:3}(e). 
By contrast, parametric oscillations involving only C supermodes are suppressed, since the C supermode inherits the normal dispersion of the uncoupled resonators. 

In \cref{fig:4}(a), we present the individual excitation of OPO 1. A zoom into the signal and idler frequencies (ii. and iii.) shows that at $\mu = 5$ and $-5$, the S and AS supermode are dominantly excited, respectively, in excellent agreement with the coupled-mode theory predictions (\cref{fig:3}(d), see Supplemental Material S2). In this configuration, the signal and idler fields are encoded with different frequencies and spatial profiles, providing an interesting platform for hyperentanglement protocols.

In \cref{fig:4}(b), the generation of OPO 2 is displayed, highlighting the evolution of the signal and idler fields. In this case, additional peaks are observed beyond the characteristic triplet of the three-ring resonator, hindering a clear identification of the excited supermodes and pointing to the presence of AMX at those frequencies.
Nevertheless, the accurate match between predicted and observed frequencies corroborates that the AS supermode is the dominant mode being excited.

Alongside the multi-FSR OPO lines, the inset of \cref{fig:3}(e) reveals excitation of the S and AS modes at $\mu = 0$, which stems from an intra-FSR OPO enabled by vertical phase-matching (\cref{fig:2}(b)).
This process corresponds to a \emph{vertical Type-II-like interband OPO} and can be regarded as a limiting case of  the diagonal Type-II-like interband OPO (OPO~1), in which the signal and idler are generated at $\mu = 0$.
Its appearance only when OPO~1 or OPO~2 are active supports the interpretation of a lower gain.
This observation is consistent with the gain profiles in \cref{fig:3}(d), which identify OPO~1 and OPO~2 as the dominant finite-$\mu$ oscillation channels.
In the following, we introduce a strategy to selectively isolate this vertical Type-II-like process.

\section{Competition suppression}\label{sec:competition_suppression}

\begin{figure}[h]
\centering
\includegraphics[width=\textwidth]{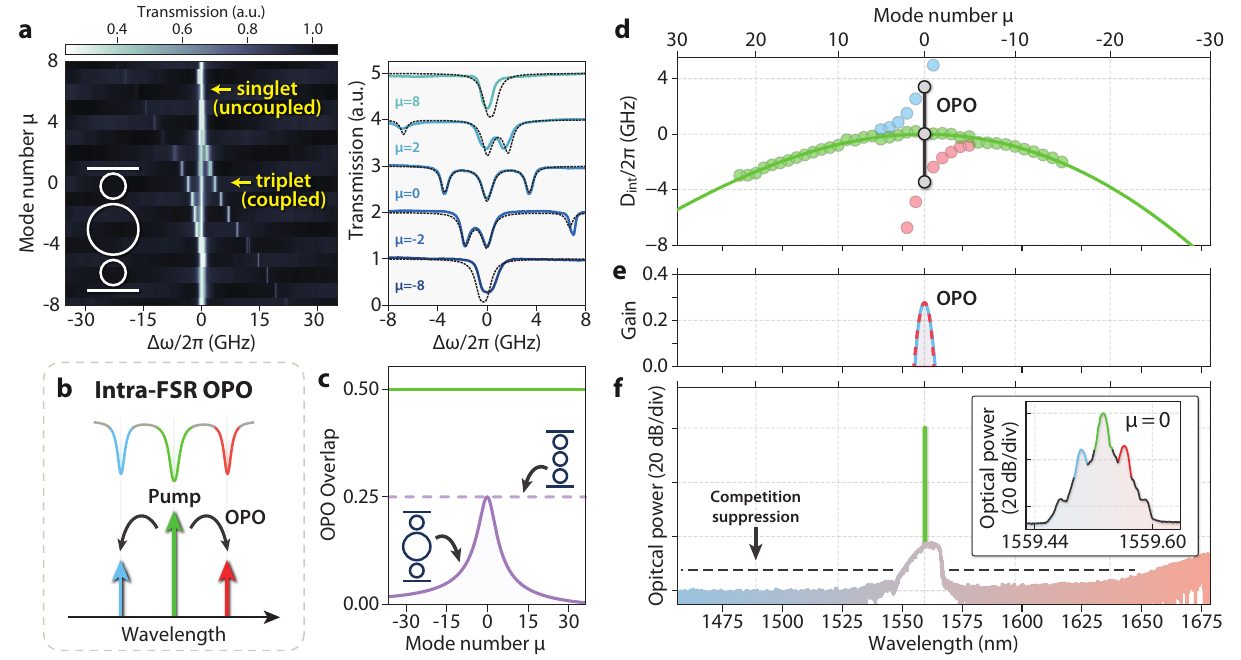}
\caption{\label{fig:5}
Intra-FSR OPO in the $|$oOo$|$ coupled resonator system.
\textbf{(a)} Left: Transmission map as a function of frequency offset ($\Delta\omega/2\pi$) and relative mode number $\mu$. 
As the wavelength changes, the different dispersion between the larger and smaller rings results in an AMX, modifying the coupling from hybridized triplets at $\mu = 0$ to uncoupled singlets at distant $\mu$.
Right: Transmission spectra for selected $\mu$ values.
Dashed lines represent the theoretical transmission.
\textbf{(b)} Schematic of the intra-FSR OPO that takes place within a single triplet. 
Pumping the C supermode results in signal and idler excitation at the AS and S modes.
\textbf{(c)} Mode overlap for the OPO process while pumping the C supermode at $\mu=0$. Purple lines assume the signal-idler excitation at AS and S supermodes (solid for $|$oOo$|$ and dashed for $|$ooo$|$). Green is the self-overlap of the C supermode.
\textbf{(d)} Integrated dispersion of the supermodes, referenced to the pumped C mode at $\mu = 0$. 
Colored markers represent experimental data, while solid curves show polynomial fits. 
Modes participating in the intra-FSR OPO at $\mu = 0$ are highlighted, corresponding to vertical phase-matching.
\textbf{(e)} Calculated parametric gain for the parametric oscillator, which peaks at $\mu = 0$.
\textbf{(f)} Optical spectrum of the competition-free intra-FSR OPO.
Inset: zoom at $\mu = 0$, showing the signal and idler sidebands.
}
\end{figure}

To further study emergent hybridized interactions, we introduce an asymmetric three-ring design ($|$oOo$|$, \cref{fig:1}(c,d)). This setup enables selective excitation of nonlinear processes while suppressing competing pathways, so that only the target OPO is active.
The device consists of a central ring with nearly twice the radius of the outer rings, creating an FSR mismatch due to differences in group velocity dispersion.
As a result, the coupling becomes wavelength dependent, with hybridization optimized only near a specific azimuthal mode number, indexed by $\mu = 0$.

\Cref{fig:5}(a) shows the transmission spectrum of this device.
By tuning integrated microheaters, we thermally align the resonances at \qty{1559.5}{\nano\meter}, ensuring the system operates in the canonical (unperturbed) hybridized regime (see Supplemental Material S1).
Away from this mode, the coupling progressively weakens due to dispersion mismatch, and the triplet structure is modified in an AMX pattern~\cite{xue2015normal}.
Around $|\mu| = 8$, the spectrum collapses into a singlet, signaling the end of strong coupling and that the central ring is no longer resonant.
Although rings 1 and 3 remain degenerate, their weak coupling (below the sideband resolution) fails to produce appreciable hybridization. 
This non-uniform spectral hybridization contrasts with the case of symmetric $|$ooo$|$ design (\cref{fig:3}(a)), where coupling persists across the spectrum.

We find that this evolving eigenmode system forms a rich platform for tailoring nonlinear interactions.
Singly pumping the C supermode triggers a vertical Type-II-like interband OPO confined to the triplet at $\mu = 0$, generating sidebands in the AS and S modes (\cref{fig:5}(b)).
In the asymmetric $|$oOo$|$ system, the spatial overlap between the modes participating in the OPO decays quickly with increasing $|\mu|$ (\cref{fig:5}(c), solid purple).
This localization suppresses competing oscillations at distant modes, in contrast with the symmetric $|$ooo$|$ configuration (dashed purple), where the supermode eigenvector structure remains unchanged across $\mu$. 
Although the C supermode retains strong self-overlap (green), FWM interactions involving only this mode are still hindered by its inherited normal dispersion.

Spectral selectivity is further supported by the dispersive band structure.
\cref{fig:5}(d) shows that an AMX effectively suppresses non-vertical phase-matching pathways, eliminating excitations at higher $|\mu|$.
This effect prohibits parametric processes such as the horizontal Type-I-like hybridized-band OPO (OPO~2), which rely on distant interaction between C and AS modes.

\cref{fig:5}(f) shows the competition-free excitation of a intra-FSR OPO at $\mu = 0$.
A conversion efficiency of \qty{-15}{\deci\bel} is observed, while delivering a broadband spectrum free of comb lines, confirming the complete suppression of competing channels.
Theoretical calculations (\cref{fig:5}(e)) show that the system supports a single parametric gain peak centered at $\mu = 0$, consistent with experimental observation.
This isolation is desired for probing parametric dynamics~\cite{inoue2019influence} and achieving high-purity nonclassical states of light~\cite{zhang2021squeezed, ulanov2025quadrature, sabattoli2021suppression, seifoory2022degenerate}.

Although singly-resonant Kerr OPOs have been examined in the past~\cite{sayson2019octave, dutt2015chip, pidgayko2023voltage}, reported signal–idler separations typically range from hundreds of \unit{\giga\hertz} to several \unit{\tera\hertz}.
In contrast, our design operates in a compact 7~GHz spacing, well-suited for commercial photodetectors, facilitating radio-frequency operation of the parametric oscillator.
This allows for simultaneous detection of the signal and idler sidebands, enabling real-time monitoring of nonclassical correlations~\cite{chen2005two, pang2025versatile} and bright squeezing dynamics~\cite{dutt2015chip, shen2025strong}.
Moreover, external coupling engineering can be implemented for efficient pump rejection~\cite{zeng2014design}.

\section{Conclusions}

In summary, we have investigated OPOs that operate across the hybridized supermodes of three-microring systems.
The coupling-induced spectral splitting gives rise to a dispersive band structure supporting multiple phase-matching topologies, including diagonal and vertical Type-II-like interband OPOs and horizontal Type-I-like hybridized-band OPOs. We emphasize that this classification is not specific to three-ring systems, but applies generally to hybridized multimode Kerr platforms with engineered dispersion and coupling.
We developed an analytical framework that accurately describes hyperparametric interband interactions within the supermode picture. 
Through combined theory and experiment, we identified the conditions for OPO onset and derived closed-form expressions for the azimuthal order of the primary sidebands.
While the present study concentrates on single-pump excitation, the underlying treatment is not restricted to this configuration and is expected to remain valid under dual-pump or more elaborate driving schemes~\cite{danilin2026intraresonance, vorobyev2025optimization, trinchao2026color}, where additional interband pathways may emerge.
We further introduced an asymmetric resonator design in which dispersion and mode overlap are co-engineered to achieve a mode-competition-free vertical Type-II-like interband OPO at ~\qty{7}{\giga\hertz}.
This work establishes practical design principles for efficient nonlinear operation in coupled-resonator platforms, while providing clear strategies to suppress undesired interactions, with applications for quantum light sources with high spectral purity~\cite{reimer2016generation, zhang2021squeezed}.

\begin{backmatter}
\bmsection{Funding}
This work was supported by São Paulo Research Foundation (FAPESP) through grants 
18/15577-5, 
18/15580-6, 
18/25339-4, 
21/10334-0, 
23/09412-1, 
24/15935-0, 
25/04049-1, 
20/04686-8, 
22/06267-8, 
24/02289-2, 
24/14425-8, 
18/21311-8, 
24/04845-0  
25/15127-3 
25/20846-9, 
25/10683-5, 
and Coordenação de Aperfeiçoamento de Pessoal de Nível Superior - Brasil (CAPES) (Finance Code 001).


\bmsection{Disclosures}
The authors declare no conflicts of interest

\bmsection{Data availability}
Data underlying the results of this paper will be made available in Zenodo upon publication (DOI to be provided).

\bmsection{Supplemental document}
See the Supplementary Material for a description of the three-ring resonator system and the analytical model of the hybridized-band OPOs presented in the main text.

\end{backmatter}


\bibliography{bib}






\end{document}


\maketitle
\tableofcontents

\section*{Table summary}
\cref{tab:OPO} summarizes the parameters used throughout the theoretical and numerical analyses.
Device-specific parameters were extracted or approximated from experimental data.

\begin{table}[h]
\centering

\setlength{\tabcolsep}{8pt}
\renewcommand{\arraystretch}{1.2}

\begin{tabular}{|c|l|c|c|c|}
\hline
\textbf{Design} & \textbf{Parameter description} & \textbf{Symbol} & \textbf{Value/Expression} & \textbf{Units} \\
\hline
\multirow{11}{*}{\makecell{\textbf{Symmetric} \\ \textbf{($|$ooo$|$)}}} 
  & First-order dispersion (uncoupled) & $D_1^{(0)}/2\pi$ & 454.53 & GHz \\
  & Second-order dispersion (uncoupled) & $D_2^{(0)}/2\pi$ & -15.33 & MHz \\
  & Third-order dispersion (uncoupled) & $D_3^{(0)}/2\pi$ & 920 & kHz \\
  & Intrinsic loss & $\kappa_i/2\pi$ & 0.75 & GHz \\
  & Extrinsic loss & $\kappa_e/2\pi$ & 0.25 & GHz \\
  & Dispersive coupling rate & $J(\mu)/2\pi$ & $J_0 + J_1 \mu +J_2 \frac{\mu^2}{2}$ & GHz \\
  & Coupling zeroth-order component & $J_0$ & $2.85$ & GHz \\
  & Coupling first-order component & $J_1$ & $-49.19$ & MHz \\
  & Coupling second-order component & $J_2$ & $840$ & kHz \\
  & Normalized pump power & $\bar{P}$ & 2.66 & GHz/m$^3$ \\
  & Laser offset & $\Delta_p/2\pi$ & 0.7 & GHz \\
\hline\hline
\multirow{9}{*}{\makecell{\textbf{Asymmetric} \\ \textbf{($|$oOo$|$)}}} 
  & First-order dispersion (small rings) & $D_1^{(\text{o})}/2\pi$ & 454.66 & GHz \\
  & Second-order dispersion (small rings) & $D_2^{(\text{o})}/2\pi$ & -15.33 & MHz \\
  & First-order dispersion (large ring) & $D_1^{(\text{O})}$ & $0.9945/2 \times D_1^{(\text{o})}$ & GHz \\
  & Second-order dispersion (large ring) & $D_2^{(\text{O})}$ & $D_2^{(\text{o})}$ & MHz \\
  & Intrinsic loss & $\kappa_i/2\pi$ & 0.75 & GHz \\
  & Coupling loss & $\kappa_e/2\pi$ & 0.25 & GHz \\
  & Coupling rate & $J/2\pi$ & $3.43/\sqrt{2}$ & GHz \\
  & Normalized pump power & $\bar{P}$ & 2.36 & GHz/m$^3$ \\
  & Laser offset & $\Delta_p/2\pi$ & 1.16 & GHz \\
\hline
\end{tabular}
\caption{Parameters used in the parametric gain simulation of OPOs in the $|$ooo$|$ and $|$oOo$|$ coupled resonator designs.}
\label{tab:OPO}
\end{table}

\section{3-ring coupled resonator}

\subsection{Principle of operation of the symmetric design}
\begin{figure}[b]
\centering
\includegraphics[width=.85\textwidth]{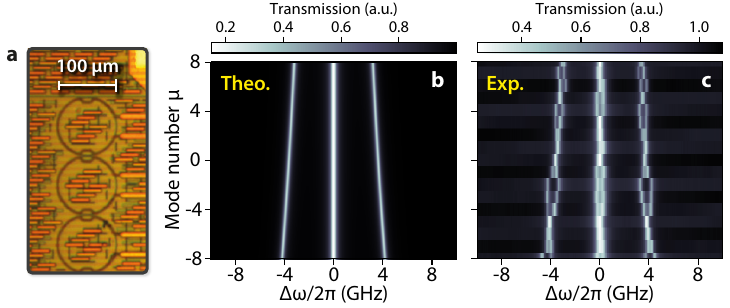}
\caption{\label{fig:ooo_transmission}
\textbf{(a)} Micrograph of the SiN $|$ooo$|$ device. \textbf{(b)} Theoretical and \textbf{(c)} experimental transmission of the device as a function of $\mu$. Increasing frequency splittings are seen due to dispersive coupling.
}
\end{figure}

Here, we describe the mode hybridization of a photonic molecule formed by three identical rings, shown in \cref{fig:ooo_transmission}(a). For low power, Kerr and thermal-induced phase-shifts can be neglected, and the coupled-mode equations for the optical field of the three resonators can be written as~\cite{tomazio2024tunable, ghosh2024controlled, zeng2014design}:
\begin{subequations}
	\begin{align}
    \frac{\partial a_{1}}{\partial \tau} &= - \left(\frac{\kappa_i+\kappa_e}{2} + i\Delta\right) a_{1} + i J a_2 + \sqrt{\kappa_e}s_{in},   \\
    \frac{\partial a_{2}}{\partial\tau} &= - \left(\frac{\kappa_i}{2} + i\Delta\right) a_{2} + i J (a_1 + a_3),   \\
    \frac{\partial a_{3}}{\partial\tau} &= - \left(\frac{\kappa_i+\kappa_e}{2} + i\Delta\right) a_{3} + i J a_2,
	\end{align}
    \label{eq:CMT}
\end{subequations}
where $a_j$ is the slow-varying complex amplitude of the $j$-ring ($j = \{1,2,3\}$), $J$ is the inter-ring linear coupling rate, $\kappa_i$ and $\kappa_e$ are the intrinsic and extrinsic losses, $\Delta = \omega - \omega_0$ is the detuning calculated from the bare-ring resonance frequency $\omega_0$, and $s_{in}$ in the input pump driving the first ring (closer to the bus waveguide).

The mode hybridization across the spectrum can be studied by examining the coupling matrix~\cite{tomazio2024tunable}:
\begin{equation}
    \mathbb{H(\mu)} =
    \begin{pmatrix}
        \omega_\mu & -J(\mu) & 0 \\
        -J(\mu) & \omega_\mu & -J(\mu) \\
        0 & -J(\mu) & \omega_\mu
    \end{pmatrix}
    .
    \label{eq:Hmatrix}
\end{equation}
Here, $\mu$ indexes the azimuthal modes of the bare ring resonator, whose optical frequency is given by $\omega_\mu = \omega_0 + D_1^{(0)} \mu + D_{\mathrm{int}}^{(0)}(\mu)$. The dispersive coefficient $J(\mu)$ appears due to wavelength-dependent overlap between the resonators.

\begin{figure}
\centering
\includegraphics[width=.5\textwidth]{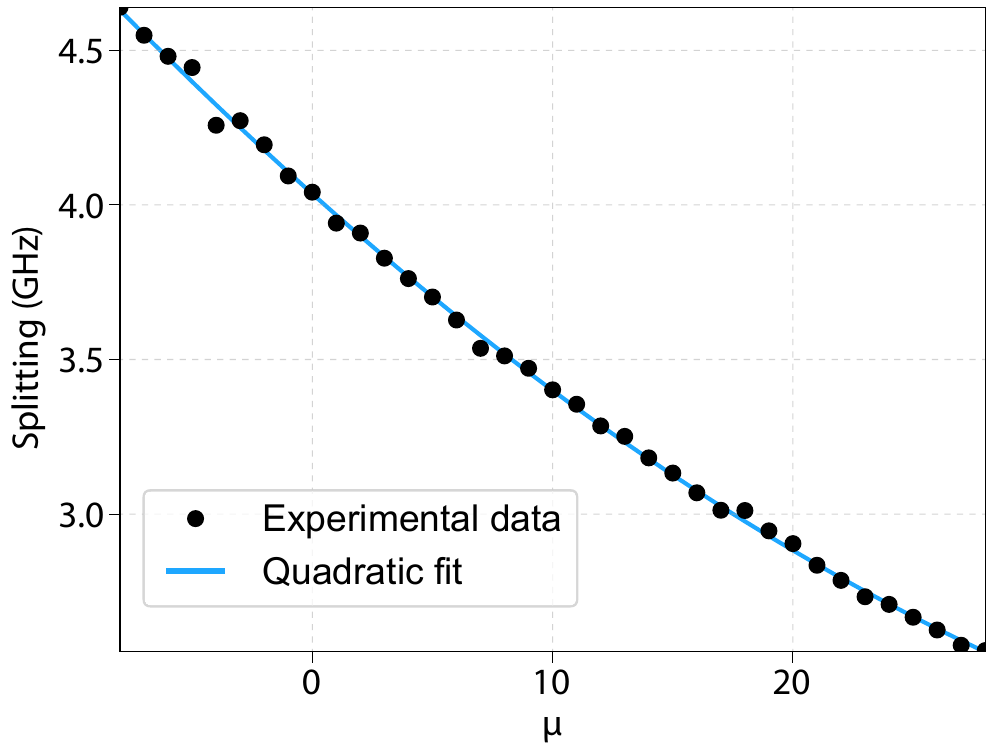}
\caption{\label{fig:splitting}
Frequency splitting as a function of the relative azimuthal mode number $\mu$. Black markers denote the experimentally measured splittings, while the blue curve shows a quadratic fit to the data (fit parameters reported in \cref{tab:OPO}).
}\
\end{figure}

Solving the eigenvalues of $\mathbb{H}(\mu)$ gives the supermodes' eigenfrequencies:
\begin{subequations}
    \begin{align}
    \omega^{(\mathrm{S})}(\mu) &= \omega_\mu - \sqrt{2}J(\mu) \nonumber \\
                               &= \omega_0 + D_1^{(0)} \mu + D^{(0)}_{int}(\mu)  - \sqrt{2}J(\mu), \\
    \omega^{(\mathrm{C})}(\mu) &= \omega_\mu \nonumber \\
                               &= \omega_0 + D_1^{(0)} \mu + D^{(0)}_{int}(\mu), \\
    \omega^{(\mathrm{AS})}(\mu) &= \omega_\mu + \sqrt{2}J(\mu) \nonumber \\
                                &= \omega_0 + D_1^{(0)} \mu + D^{(0)}_{int}(\mu) + \sqrt{2}J(\mu),
    \end{align}
    \label{eq:couplingdispersion}
\end{subequations}
corresponding to the symmetric (S), central (C), and anti-symmetric (AS) supermodes, with $\omega_\mathrm{S}<\omega_\mathrm{C}<\omega_\mathrm{AS}$, as shown in Fig. 3(a) of the main text.
\cref{eq:couplingdispersion} shows that the S and AS modes experience additional dispersion due to coupling:
\begin{subequations}
\begin{align}
    &D^{(\mathrm{S})}_{int}(\mu) \equiv D^{(0)}_{int}(\mu)  - \sqrt{2}J(\mu), \\
    &D^{(\mathrm{AS})}_{int}(\mu) \equiv D^{(0)}_{int}(\mu)  + \sqrt{2}J(\mu),
\end{align}
\end{subequations}
while the C supermode retains the bare-ring dispersion:
\begin{equation}
    D^{(\mathrm{C})}_{int}(\mu)= D^{(0)}_{int}(\mu).
\end{equation}
In addition, the supermodes S and AS are equally spaced from C by $\sqrt{2}J(\mu)$, resulting in a frequency splitting that increases with the wavelength, as shown in \cref{fig:ooo_transmission}(b,c). \cref{fig:splitting} shows the measured frequency splitting as a function of the relative azimuthal mode order $\mu$. A quadratic fit provides an accurate description.

In the spatial domain, the eigenvectors of $\mathbb{H}(\mu)$ describe the supermode field distributions, illustrated in Fig. 1(b) of the main text. 
We express each supermode profile as projected onto the bare-ring (uncoupled) basis as:
\begin{subequations}
\begin{align}
    &\vec{b}_\mathrm{S}  = \frac{1}{2} \, \left(1 ~ \sqrt{2} ~ 1 \right)^T, \\
    &\vec{b}_\mathrm{C}  = \frac{1}{\sqrt{2}} \, \left(-1 ~ 0 ~ 1\right)^T, \\
    &\vec{b}_\mathrm{AS} = \frac{1}{2} \, \left(1 ~ -\sqrt{2} ~ 1 \right)^T,
\end{align}
\label{eq:eigenvectors}
\end{subequations}
where the $j$-th element of each eigenvector corresponds to the electric field amplitude of the supermode in ring $j$.
For identical rings, these eigenvectors are invariant with $\mu$, forming a fixed (canonical) basis.

\subsection{Principle of operation of the asymmetric design}

\begin{figure}
\centering
\includegraphics[width=.85\textwidth]{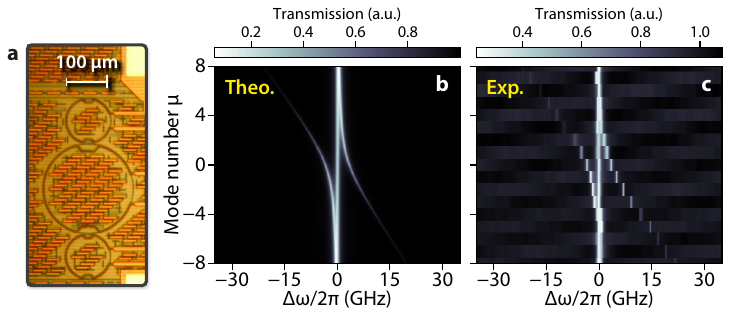}
\caption{\label{fig:oOo_transmission}
\textbf{(a)} Micrograph of the $|$oOo$|$ device. 
\textbf{(b)} Theoretical and \textbf{(c)} experimental transmission maps of the device, showing the evolution of avoided crossings.
}
\end{figure}

The asymmetric three-ring design, denoted as $|$oOo$|$, comprises a central ring whose radius is approximately twice that of the two outer rings (\cref{fig:oOo_transmission}(a)). 
This results in distinct FSRs between the central and outer rings, leading to avoided mode crossings in the transmission spectrum, as shown in \cref{fig:oOo_transmission}(b,c). These avoided crossings can be dynamically tuned using integrated microheaters~\cite{tomazio2024tunable}.

The coupling matrix for the $|$oOo$|$ deisgn reads:
\begin{equation}
    \mathbb{H(\mu)} =
    \begin{pmatrix}
        \omega^{(\text{o})}_\mu & -J(\mu) & 0 \\
        -J(\mu) & \omega^{(\text{O})}_\mu & -J(\mu) \\
        0 & -J(\mu) & \omega^{(\text{o})}_\mu
    \end{pmatrix}
    ,
\end{equation}
where $\omega^{(\text{o})}_\mu$ and $\omega^{(\text{O})}_\mu$ are the resonance frequencies of the outer (smaller) and central (larger) rings, respectively.
Because the FSR of the larger ring is approximately half that of the smaller rings, coupling occurs at every mode $\mu$ of the smaller outer rings, and every second mode of the larger central ring. For consistency, we index all modes using $\mu$, corresponding to the azimuthal mode number of the outer rings where hybridization occurs.

\begin{figure}[b]
\centering
\includegraphics[width=1\textwidth]{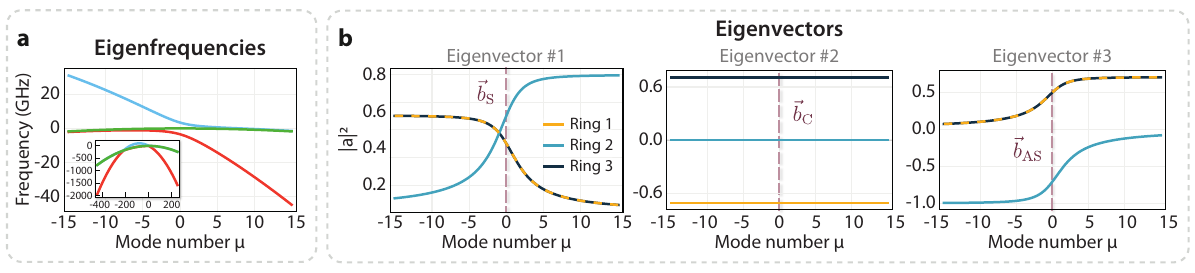}
\caption{\label{fig:SM_eigenproblem}
\textbf{(a)} Eigenfrequencies and \textbf{(b)} eigenvectors of the $|$oOo$|$ three-ring design as a function of the mode number $\mu$. 
The avoided mode crossing in (a) arises from the intersection of two dispersion parabolas (inset), which correspond to the individual dispersion of the smaller and larger rings. 
At $\mu=0$, both eigenvalues and eigenvectors restore the canonical coupled basis $\{\vec{b}_\mathrm{S}, \vec{b}_\mathrm{C}, \vec{b}_\mathrm{AS}\}$. 
}\
\end{figure}

Solving the eigenvectors and eigenfrequencies of $\mathbb{H(\mu)}$ reveals the mechanisms of the avoided crossing interaction (\cref{fig:SM_eigenproblem}).
Near a mode index $\mu = 0$, the resonances of all three rings are nearly degenerate, and the system behaves similarly to the symmetric $|$ooo$|$ design, recovering the canonical hybridized basis $\{\vec{b}_\mathrm{S}, \vec{b}_\mathrm{C}, \vec{b}_\mathrm{AS}\}$. 
However, as $|\mu|$ increases, the dispersion difference between the small and large rings lifts this degeneracy.
The increasing spectral mismatch gradually weakens the hybridization, and the system transitions toward a decoupled regime. In this limit, the eigenmodes approach the normal modes of two independent subsystems:
\begin{enumerate}
    \item The weakly coupled outer rings (1 and 3) resulting in a symmetric and anti-symmetric doublet:
    \begin{align}
	&\vec{c}_\mathrm{S}(\mu\to\infty) = \frac{1}{\sqrt{2}}(1~0~1)^T,   \\
	&\vec{c}_{\mathrm{AS}}(\mu \to\infty) = \frac{1}{\sqrt{2}}(-1~0~1)^T,   \\
    \end{align}
    \item The singlet of the central ring:
    \begin{align}
	&\vec{a}_\mathrm{2}(\mu\to\infty) = (0~1~0)^T.  
    \end{align}
\end{enumerate}
Since the outer rings are weakly coupled (i.e., in the sideband unresolved regime) and the central ring is not directly driven by the bus waveguide, the spectrum exhibits only singlet resonances at large values of $|\mu|$, as shown in \cref{fig:oOo_transmission}.
This change enables co-engineering of both eigenfrequencies (dispersion) and eigenvectors (mode overlap), which we exploited to suppress parasitic FWM interactions during the OPO operation shown in Fig.~5(f) of the main text. More details are given in \cref{sec:SM_OPO_model}.

\section{Model for OPO in three coupled rings}\label{sec:SM_OPO_model}

In our identical three-ring coupled system ($|$ooo$|$), whose bare rings exhibit normal dispersion, two independent mechanisms for parametric oscillation have been identified, referred to as OPO~1 (diagonal Type-II-like interband OPO) and OPO~2 (horizontal Type-I-like hybridized-band OPO) (see Fig.~3(b) in the main text).

OPO~1 arises when a single pump is tuned into the C supermode at $\mu=0$, leading to the excitation of an AS supermode at $\mu = -5$ and an S supermode at $\mu=5$.
On the other hand, OPO~2 emerges at AS supermodes at $\mu = \pm 21$.
Below, we detail the calculation of the parametric gain of these processes, presented in Fig. 3(d) of the main text.
\cref{tab:OPO} presents the simulation parameters used.

The time evolution for the signal and idler at the onset of OPO 1 can be described within a nonlinear coupled-mode equation framework~\cite{herr2012universal}:
\begin{align}
    \frac{da^{(\mathrm{S})}_\mu}{dt} &= -\left(\frac{\kappa_{\mathrm{S}}}{2}+i\Delta^{(\mathrm{S})}_\mu \right)a^{(\mathrm{S})}_\mu + i2g_{\mathrm{S}} \Gamma_{\mathrm{S,C}} \left|a^{(\mathrm{C})}_0\right|^2 a^{(\mathrm{S})}_\mu + ig_{\mathrm{S}}\Lambda_{\mathrm{S}}^{(\mathrm{C}, \mathrm{AS}, \mathrm{C})} \left(a^{(\mathrm{C})}_0\right)^2 \left(a^{(\mathrm{AS})}_{-\mu}\right)^*, \\
    \frac{d\left(a^{(\mathrm{AS})}_{-\mu}\right)^*}{dt} &= -\left(\frac{\kappa_{\mathrm{AS}}}{2}-i\Delta^{(\mathrm{AS})}_{-\mu} \right)\left(a^{(\mathrm{AS})}_{-\mu}\right)^* - i2 g_\mathrm{AS} \Gamma_{\mathrm{AS,C}} \left|a^{(\mathrm{C})}_0\right|^2 \left(a^{(\mathrm{AS})}_{-\mu}\right)^* - ig_\mathrm{AS} \Lambda_{\mathrm{AS}}^{(\mathrm{C}, \mathrm{S}, \mathrm{C})} {\left(a^{(\mathrm{C})}_0\right)^*}^2 a^{(\mathrm{S})}_\mu.
\label{eq:OPOs}
\end{align}

Here, $j = \{\mathrm{S},\mathrm{C},\mathrm{AS}\}$ denote the supermodes. $a_j$ are the complex amplitudes, $\kappa_j$ are the total loss rates, and
\begin{align}
    \Delta_\mu^{j} &= (\omega_\mu^{(j)}  -\omega_0^{(\mathrm{C})}-D_1^{(\mathrm{C})}\mu) + \Delta_p\\
    &=\Delta_p + D^{(j)}_{int}(\mu)
\end{align}
is the detuning with $\Delta_p=\omega^{(\mathrm{C})}_0-\omega_p$ being the pump detuning with respect to the laser frequency $\omega_p$. 
$g_j$ denotes the Kerr coefficients and the overlap integrals $\Gamma_{j,k}$ (XPM) and $\Lambda_{\eta}^{\alpha,\beta,\gamma}$ (FWM) follow Refs.~\cite{tomazio2024tunable,trinchão2025mapping}:
\begin{align}
    \Gamma_{j,k} &= \frac{1}{4}\, \bigintssss_V \epsilon^2
                    \left| \vec{\Upsilon}_k^* \cdot \vec{\Upsilon}_j \right|^2\, dV,
                    \label{eq:XPM} \\
    \Lambda_\eta^{\alpha,\beta,\gamma} &= \frac{1}{4}\, \bigintssss_V \epsilon^2
                    \left( \vec{\Upsilon}_\eta^* \cdot \vec{\Upsilon}_\mu \right)
                    \left( \vec{\Upsilon}_\beta^* \cdot \vec{\Upsilon}_\alpha \right)\, dV,
                    \label{eq:FWM}
\end{align}
where $\vec{\Upsilon}_j$ denotes the optical field profile of the $j$-supermode and $V$ the total volume of the three-ring system.
Evaluation \cref{eq:XPM,eq:FWM} is greatly simplified by decomposing the supermode profiles ${\vec{\Upsilon}j}$ into the eigenvector description ${\vec{b}j}$, following Ref.~\cite{tomazio2024tunable}.
Moreover, because the symmetric $|$ooo$|$ system features a $\mu$-independent eigenbasis, all $\Gamma_{j,k}$ and $\Lambda^{\alpha, \beta, \gamma}_{\eta}$ are likewise independent of $\mu$.

Under the undepleted pump approximation~\cite{herr2012universal}, \cref{eq:OPOs} forms an eigenvalue problem for $a^{(\mathrm{S})}_\mu$ and $\left(a^{(\mathrm{AS})}_{-\mu}\right)^*$, whose eigenvalues correspond to the parametric gain:
\begin{align}
    g^{(\mathrm{S})}(\mu) &= - \frac{\kappa + i \left(\delta_\mathrm{S}(\mu) + \delta_\mathrm{AS}(-\mu)\right)}{2} + \sqrt{\left(\Lambda \bar{P} \right)^2 - \left(\frac{\delta_\mathrm{S}(\mu) - \delta_\mathrm{AS}(-\mu)}{2} \right)^2}, \label{eq:gainAS} \\
    g^{(\mathrm{AS})}(\mu) &= - \frac{\kappa + i \left(\delta_\mathrm{S}(-\mu) + \delta_\mathrm{AS}(\mu)\right)}{2} + \sqrt{\left(\Lambda \bar{P} \right)^2 - \left(\frac{\delta_\mathrm{S}(-\mu) - \delta_\mathrm{AS}(\mu)}{2} \right)^2}.
    \label{eq:gainS}
\end{align}
(The expression for $g^{(\mathrm{AS})}$ follows from the analogous eigenproblem involving $a^{(\mathrm{AS})}_\mu$ and $\left(a^{(\mathrm{S})}_{-\mu}\right)^*$). We describe all terms below.

Here, $\bar{P} = \frac{\omega_\mu c n_2}{n_0 n_g} |a^{(\mathrm{C})}_0|^2$ is the normalized intracavity pump power with units of $[\mathrm{Hz}/\mathrm{m}^3]$ and the frequency-mismatch terms read $\delta_j = \Delta_\mu^{(j)}-2\Gamma_{j,\mathrm{C}}\bar{P}$.
For the S and AS branches, these become:
\begin{align}
    \delta_\mu^{(\mathrm{S})}(\mu) &=  D^{(0)}_{int}(\mu) - \sqrt{2}J(\mu) + \Delta_p - 2\Gamma_{\mathrm{S},\mathrm{C}}\bar{P}, 
    \label{eq:freq_mis1}\\
    \delta_\mu^{(\mathrm{AS})}(\mu) &=  D^{(0)}_{int}(\mu) + \sqrt{2}J(\mu) + \Delta_p - 2\Gamma_{\mathrm{AS},\mathrm{C}}\bar{P}.
    \label{eq:freq_mis2}
\end{align}
\cref{eq:freq_mis1,eq:freq_mis2} show that parametric gain in the coupled-cavity system is constrained by the simultaneous balance of Kerr-induced XPM, the bare-ring group-velocity dispersion, the inter-ring coupling, and the pump detuning.
Moreover, \cref{eq:gainS,eq:gainAS} takes advantage of the symmetry of the S and AS modes~\cite{trinchão2025mapping} to further streamline the notation:
\begin{align}
    \kappa&=\kappa_\mathrm{S}=\kappa_\mathrm{AS} \\
    \Gamma &=  \Gamma_{\mathrm{S,C}} =\Gamma_{\mathrm{AS,C}} = \frac{1}{4},\\
    \Lambda &= \Lambda_{\mathrm{S}}^{(\mathrm{C}, \mathrm{AS}, \mathrm{C})}=\Lambda_{\mathrm{AS}}^{(\mathrm{C}, \mathrm{S}, \mathrm{C})} = \frac{1}{4}.
\end{align}

Parametric gain analysis is facilitated by expanding the inter-ring coupling function $J(\mu)$ up to second order in the azimuthal mode index, which provides an accurate description in the relevant spectral region (see \cref{fig:splitting}):
\begin{equation}
    J(\mu) = J_0 + J_1 \mu +J_2 \frac{\mu^2}{2}.
\end{equation}

The primary comb lines associated with the OPO I process are determined by identifying the azimuthal order $\mu$ that maximizes the parametric gain. This condition is obtained by imposing
\begin{equation}
    \frac{d\,\mathrm{Re}( g^{(\mathrm{S})}(\mu))}{d\mu} = 0\quad \text{or} \quad  \frac{d \,\mathrm{Re}(g^{(\mathrm{AS}))}(\mu)}{d\mu} = 0,
\end{equation}
which yields a closed-form expression:
\begin{equation}
    \boxed{\mu_{th,I}=\frac{\sqrt{2}J_1}{D_2^{(0)}}}.
    \label{eq:muthI}
\end{equation}

\begin{figure}
\centering
\includegraphics[width=1\textwidth]{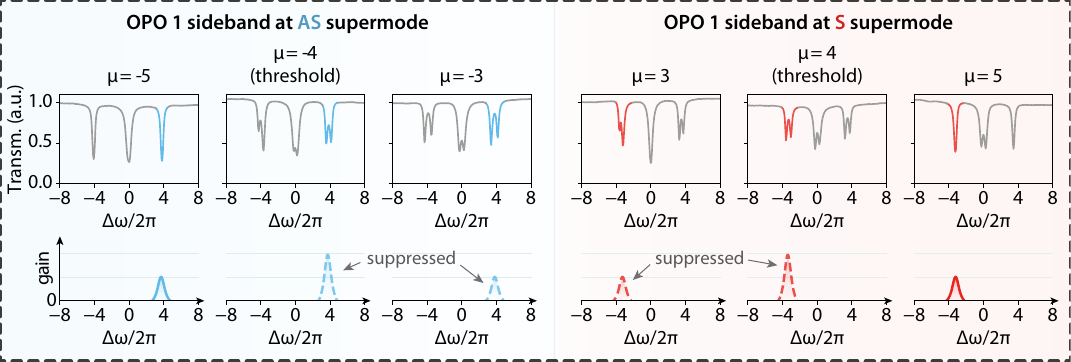}
\caption{\label{fig:thresholdsup}
In the ideal case, the maximum parametric gain for the OPO~1 is centered at $\mu_{th,I}\approx\pm4$, with neighboring azimuthal modes experiencing comparable gain. Experimentally (top panels), backscattering-induced mode splitting at $\mu=\pm4$ and $\mu=\pm3$ locally perturbs the dispersion and suppresses the gain at these modes. As a result, the effective maximum gain shifts to the neighboring modes at $\mu=\pm5$ (bottom panels, schematics).
}\
\end{figure}

\eqref{eq:muthI} shows that the position of the maximum-gain lobe is set solely by the bare-ring dispersion and the odd component of the coupling function $J(\mu)$. As a result, the inter-ring coupling plays a role comparable to that of dispersion in determining the nonlinear dynamics and frequency-comb formation in coupled resonators, and must be explicitly considered during device design.
For the present device, we obtain $\mu_{th,I}\approx4$, corresponding to the center of the gain lobe predicted by the model.
In practice, however, spurious back-scattering between CW and CCW modes introduces additional splittings at $\mu\in\{-4, -3, 3,4\}$, as shown in \cref{fig:thresholdsup}.
These perturbations locally modify the effective dispersion and suppress efficient phase matching at these modes, reducing their gain~\cite{ulanov2025quadrature}.
Nevertheless, because the gain lobe predicted by the model is sufficiently broad, neighboring modes at $\mu=\pm5$ can still reach oscillation and be populated~\cite{herr2012universal} (see \cref{fig:thresholdsup}), which is consistent with our experimental observations.

\begin{figure}
\centering
\includegraphics[width=.54\textwidth]{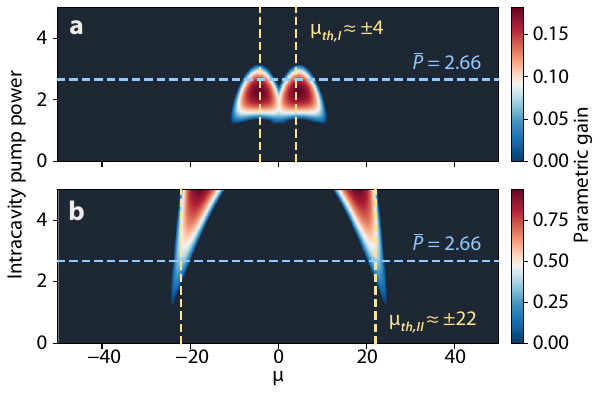}
\caption{\label{fig:symmetric_gain_map}
Parametric gain lobes versus intracavity pump power $\bar{P}$ for the symmetric coupled resonator.
Panels show the gain associated with OPO~1 (diagonal Type-II-like interband OPO) \textbf{(a)} and OPO~2 (horizontal Type-I-like hybridized-band OPO) \textbf{(b)}
 for $\Delta_p/2\pi=0.7$~GHz, together with the corresponding threshold azimuthal orders.
Horizontal lines indicate the pump power values used in Fig.~3(d) of the main text.
}\
\end{figure}

The same methodology can be applied to the OPO II process by considering the appropriate AS supermodes involved in that interaction. 
Following the same linear stability analysis and gain-maximization procedure, we obtain
\begin{equation}
    \boxed{\mu_{th,II} =
    \sqrt{\frac{\bar{P} - 2\sqrt{2}J_0 -2\Delta_p}{D_2^{(0)}+\sqrt{2}J_2}
    }}.
    \label{eq:muthII}
\end{equation}
In contrast to \eqref{eq:muthI}, the expression for $\mu_{th,II}$ depends exclusively on the \emph{even} components of the coupling function $J(\mu)$. 
For the parameters considered here, \eqref{eq:muthII} predicts $\mu_{th,II}\approx22$; however, backscattering-induced mode splitting shifts the effective maximum gain to the neighboring modes $\mu=\pm21$ (see Figs 3 and 4 of the main text).
Moreover, \cref{fig:symmetric_gain_map} presents the parametric gain of OPO I and OPO II as a function of the circulating pump power. OPO I is supported only within a relatively narrow power window, whereas OPO II persists over a broader range and becomes the dominant nonlinear interaction at higher pump levels.

\begin{figure}[b]
\centering
\includegraphics[width=.54\textwidth]{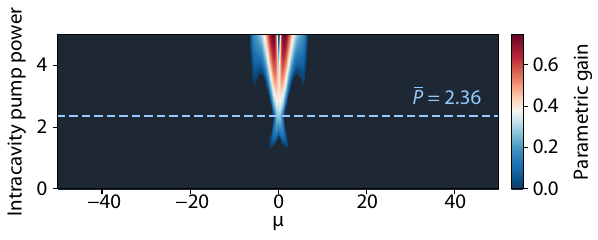}
\caption{\label{fig:asymmetric_gain_map}
Parametric gain lobes versus intracavity pump power $\bar{P}$ for the asymmetric $|$oOo$|$ coupled resonator for $\Delta_p/2\pi=1.16$~GHz. 
The horizontal line indicates the pump power value used in Fig.~5(e) of the main text.
}\
\end{figure}

For the asymmetric $|$oOo$|$ configuration, the analysis must incorporate the larger mode volume of the central resonator (approximately twice that of the outer rings) as well as the $\mu$-dependent dispersive evolution of the overlap integrals, which vary due to the evolving coupled mode basis deviating from the canonical form (see \cref{fig:SM_eigenproblem} and 5(c) in the main text).
Since the avoided mode crossing is the dominant effect on shaping the spectrum of the device, the dispersive nature of the coupling function is neglected (see \cref{tab:OPO}).
Under these considerations, the parametric gain for the intra-FSR OPO (vertical Type-II-like interband OPO) can be calculated
, as presented in Figs. 5(e). 
In \cref{fig:asymmetric_gain_map}, we show the $|$oOo$|$ configuration parametric gain as a function of the circulating power.
In this case, the gain remains effectively confined around $\mu=0$, consistent with a vertical Type-II-like interband OPO regime.

\bibliography{bib}
